\documentstyle[aps,preprint]{revtex}
\begin{document}
\draft
\preprint{\vbox{\hbox{\tt SOGANG-HEP 271/00} }}
\title{BFT embedding of noncommutative D-brane system}
\author{Soon-Tae Hong\footnote{electronic address:sthong@phya.snu.ac.kr},
  Won Tae Kim\footnote{electronic address:wtkim@ccs.sogang.ac.kr}, 
  Young-Jai Park\footnote{electronic address:yjpark@ccs.sogang.ac.kr}, and
  Myung Seok Yoon\footnote{electronic address:younms@physics.sogang.ac.kr}}
\address
{Department of Physics and Basic Science Research Institute, \\
Sogang University, Seoul 121-742, Korea }
\date{\today}
\maketitle
\begin{abstract}
We study noncommutative geometry in the framework of the 
Batalin-Fradkin-Tyutin(BFT) scheme, which converts second class constraint 
system into first class one. In an open string,
theory noncommutative geometry 
appears due to the mixed boundary conditions having second class constraints, 
which arises in string theory with $D$-branes under a constant Neveu-Schwarz 
$B$-field. Introduction of a new coordinate $y$ on $D$-brane through BFT 
analysis allows us to obtain the commutative geometry 
with the help of the first class constraints, and  
the resulting action corresponding to the first
class Hamiltonian in the BFT Hamiltonian formalism has a
new local symmetry.
\end{abstract}
\pacs{PACS numbers: 11.10., 11.10.Ef, 11.30.}

\newpage

\section{Introduction}
\label{sec:intro}
The phenomena that spacetime does not commute appear in many physical
systems. As an example, noncommutativity of spacetime arises
in quantum mechanics of a particle constrained to move on a
two-dimensional plane, interacting minimally with a constant magnetic 
field $B$ perpendicular to the plane when the velocity of
particle is small or the strength of magnetic field is large~\cite{bw,bs}. 
Also we encounter noncommutative spacetime when we
deal with $D$-branes in string theory~\cite{sw} or we consider the
context of the matrix model of $M$-theory~\cite{bfss,bilal}. 
In the matrix model the noncommutativity of spacetime is due to 
non-abelian Yang-Mills gauge theory induced from parallel $N$ $D$-branes
at coinciding positions, all on top of each other.  The noncommutativity of 
spacetime takes place in the direction perpendicular to $D$-branes. 
Furthermore, when the end points of the open strings are attached 
to $D$-branes in the presence of a constant Neveu-Schwarz (NS) $B$-field, 
the spacetime coordinates on $D$-branes also do not commute in the 
directions parallel to $D$-branes. 

On the other hand, in an open string theory the noncommutativity of 
geometry is due to the mixed boundary conditions, which are neither 
Neumann conditions nor Dirichlet ones.  Since these boundary conditions 
form second-class constraints, spacetime coordinates become noncommutative.
   Recently, in the framework of Dirac's Hamiltonian analysis for this 
constraint system, the noncommutativity of geometry has been studied in 
Ref.~\cite{ko}.  In the viewpoint of constraint algebra, some kind of local
symmetry has been broken in this model, since the constraints are 
second-class ones.  Therefore, according to the spirit of constraint 
analysis, one could conjecture that if we could recover this symmetry, 
the noncommutativity of spacetime coordinates would disappear, and 
boundary conditions would consist of a first-class constraint system.
Furthermore, when we consider the system giving noncommutative geometry, 
ordinary product between functions must be replaced by the Moyal
bracket~\cite{sw}. This bracket is of course reduced to the ordinary
product in commutative geometry. In general, the Moyal bracket has a
complex structure. If we can convert noncommutative geometry into
commutative one without changing physics, the algebra of functions
becomes simple.  In fact, the commutative geometry can be obtained by 
constructing first-class constraint system.

In this paper, in order to convert a second-class constraint system into 
a first-class one to obtain the commutative geometry, we use the potent 
method of Batalin, Fradkin, and Tyutin (BFT)~\cite{BFT}, which is compared 
with Dirac formalism.  In fact, the BFT embedding has been widely applied 
to interesting Hamiltonian systems with second class 
constraints~\cite{brr,fg}, such as Chern-Simons(CS) theory~\cite{kp,bhr}, 
chiral Schwinger model~\cite{kkppy}, chiral bosons~\cite{bw,w,an}, to yield 
the theories with first class constraints with additional 
symmetries~\cite{sh}. 

In Sec.~\ref{sec:B-field}, before we use the BFT method, we will construct 
the Hamiltonian system with second-class constraints and the nontrivial 
commutators in terms of Dirac brackets by using discretized 
Lagrangian of open string.  In Sec.~\ref{sec:BFT}, using the BFT procedure, we 
will find the modified Hamiltonian having desired canonical structure defined 
in terms of usual Poisson brackets, and the effective Lagrangian which gives 
simple commutative geometry since the constraints become first-class ones. 
Also we will obtain new symmetries recovered via the BFT embedding.  Finally, 
in Sec.~\ref{sec:dis} we will summarize the results of the BFT embedding.

\section{Open Strings Attached At $D$-branes In The Presence Of Constant
  $B$-Field}
\label{sec:B-field}
We consider the open strings whose end points are attached at
$Dp$-branes in the presence of a constant NS $B$-field. Although for 
arbitrary $Dp$-branes it is highly nontrivial to find the system 
with commutative geometry, we can systematically construct this system 
by introducing new variables through the BFT embedding. The number of 
necessary coordinates introduced by this BFT method depends on the residual 
rank, $r \leq p+1$, of the matrix $B_{ij}$ from being gauged away. 

First, let us briefly recapitulate the geometry of open string theory 
before we introduce new variables.  Our worldsheet action of the open string, 
to which background gauge fields are coupled, is given as
\begin{eqnarray}
  \label{eq:open-S}
  S &=& \frac{1}{4\pi\alpha'} \int_\Sigma d^2 \sigma \left[ g_{ij}
  \partial_a x^i \partial^a x^j + 2\pi\alpha' B_{ij}\epsilon^{ab}
  \partial_a x^i \partial_b x^j \right] \nonumber \\
  & & \quad\ + \int d\tau \left. A_i \partial_\tau x^i \right|_{\sigma=\pi} -
  \int d\tau \left. A_i \partial_\tau x^i \right|_{\sigma = 0},
\end{eqnarray}
where $\Sigma$ is the string worldsheet with the metric 
$\eta_{ab} = {\rm diag}(+, -)$. 
The metric of target space on $D$-brane is taken to be Euclidean, 
$g_{ij} = \delta_{ij}$.  Note that the action (\ref{eq:open-S}) has two 
$U(1)$ gauge symmetries. One of them is $\lambda$-symmetry for the transformation
of $A \rightarrow A + d\lambda$ and the other is $\Lambda$-symmetry for
the transformation with a form of both $B \rightarrow B + d\Lambda$ and $A
\rightarrow A + \Lambda$.  From the open string action (\ref{eq:open-S}) one can 
obtain the equations of motion to determine the mixed boundary conditions 
as follows 
\begin{equation}
  \label{eq:open-bc}
  \left. g_{ij} \partial_\sigma x^j + 2\pi\alpha' {\cal F}_{ij}
  \partial_\tau x^j \right|_{\sigma=0,\pi} = 0,
\end{equation}
where ${\cal F} = B - F$ and $F = dA$.  Without background fields $B$
and $A$, the boundary conditions (\ref{eq:open-bc}) are Neumann
boundary ones, $\partial_\sigma x^i = 0$ at $\sigma =
0,\pi$, while for $B_{ij} \rightarrow \infty$ or $g_{ij} 
\rightarrow 0$, the boundary conditions become Dirichlet, $\partial_\tau 
x^i(\tau)=0$ on $D$-branes.

Next, in order to study the bulk near boundary and boundary itself in detail, 
let us discretize the open string action and the boundary conditions along
the direction of $\sigma$ with equal spacing, $\epsilon =
\frac{\pi}{N}$ which is taken to be very small, and then the integral
for $\sigma$ is changed to the sum, namely, $\int_0^\pi d\sigma \cdots = 
\sum_{a=0}^{N} \epsilon\cdots$ where $N$ is integer.  Now, let us define 
$x_{(a)}^i$ by $x_{(a)}^i (\tau) = x^i (\tau, \sigma)|_{\sigma=a \epsilon} $. 
The discretized action of the open string action (\ref{eq:open-S}) is then 
written as 
\begin{eqnarray}
S &=& \frac{1}{4\pi\alpha'} 
   \int d\tau \sum_{a=0} 
   \left[
   \epsilon \left(\dot{x}_{(a)}^i\right)^2 - \frac{1}{\epsilon} 
   \left( x_{(a+1)}^i - x_{(a)}^i \right)^2 + 4\pi\alpha' B_{ij}
   \dot{x}_{(a)}^i \left( x_{(a+1)}^j - x_{(a)}^j \right)\right]
   \nonumber\\
  & & + \int d\tau \sum_{a=0} A_i \partial_\tau (x_{(a+1)}^i - x_{(a)}^i),  
\label{eq:d-S}
\end{eqnarray}
and the mixed boundary conditions are given by
\begin{equation}
  \label{eq:d-bc}
  g_{ij} \frac{1}{\epsilon} (x_{(1)}^j - x_{(0)}^j) 
  + 2\pi\alpha' {\cal F}_{ij} \dot{x}_{(0)}^j = 0,
\end{equation}
where we have taken only $\sigma=0$ case of the boundary conditions since we
have an interest at the boundary $\sigma = 0$ only. Note that 
$x_{(0)}^i$ denotes the end of the open strings. The action (\ref{eq:d-S})
and the conditions (\ref{eq:d-bc}) become continuous in
$\sigma$-direction in the large $N$ limit (equivalently in the small 
$\epsilon$ limit).  From the action (\ref{eq:d-S}) we obtain the canonical 
momenta of $x_{(a)}^i$
\begin{equation}
  \label{eq:d-mom}
  p_{(a)i} = \frac{\epsilon}{2\pi\alpha'} 
  \left[ g_{ij}\dot{x}_{(a)}^j + 2\pi\alpha' B_{ij} \frac{1}{\epsilon}
  \left(x_{(a+1)}^j - x_{(a)}^j \right) - 2\pi\alpha'
  \frac{1}{\epsilon} A_i \delta_{a,0} \right].
\end{equation}
The combination of the boundary conditions
(\ref{eq:d-bc}) and the canonical momenta (\ref{eq:d-mom}) gives the
primary constraints
\begin{equation}
  \label{eq:d-constr}
  \Omega_i = \frac{1}{\epsilon}\left[ (2\pi\alpha')^2 {\cal F}_{ij}
  (p_{(0)}^j + A^j) + M_{ij} (x_{(1)}^j - x_{(0)}^j )\right] \approx 0,
\end{equation}
where $M_{ij} = \left[ g- (2\pi\alpha')^2 {\cal F}g^{-1}B\right]_{ij}$. 
By taking the Legendre transformation, from the Lagrangian of the action
(\ref{eq:d-S}) we can obtain the primary Hamiltonian of the form
\begin{equation}
  \label{eq:d-Hp}
  H_p = H_c + u^i \Omega_i,
\end{equation}
where $u^i$ are the multipliers and the canonical Hamiltonian is given by 
\begin{eqnarray}
  \label{eq:d-Hc}
  H_c &=& \frac{1}{4\pi\alpha'} \frac{1}{\epsilon} \sum_{a=1}
  \left\{ (2\pi\alpha')^2 \left[ p_{(a)i}- B_{ij} (x_{(a+1)}^j -
  x_{(a)}^j)\right]^2 + (x_{(a+1)}^i - x_{(a)}^i)^2 \right\} \nonumber\\
  & & + \frac{1}{4\pi\alpha'} 
        \frac{1}{\epsilon} 
        (\theta^{-1}{\cal F}^{-1}g)_{ij} 
        (x_{(1)}^i - x_{(0)}^i)(x_{(1)}^j - x_{(0)}^j)
\end{eqnarray}
with
\begin{equation}
  \label{eq:d-theta}
  \theta_{ij}^{-1} = - \frac{1}{(2\pi\alpha')^2} \left[ (g -
  2\pi\alpha' {\cal F}) {\cal F}^{-1} (g + 2\pi\alpha' {\cal F})
  \right]_{ij}.
\end{equation}
On the other hand, the constraints (\ref{eq:d-constr}) form a second-class 
system since their Poisson brackets are given as 
\begin{eqnarray}
  \label{eq:d-PB}
  \Delta_{ij} &\equiv& \{ \Omega_i, \Omega_j \} \nonumber \\
     &=& \frac{1}{\epsilon^2}(2\pi\alpha')^2 \left[ (G + M)g^{-1}
     {\cal F}\right]_{ij},
\end{eqnarray}
where
\begin{equation}
  \label{eq:d-G}
  G_{ij} = \left[ (g + 2\pi\alpha' {\cal F}) g^{-1} (g - 2\pi\alpha'
  {\cal F}) \right]_{ij} = \left[ g - (2\pi\alpha')^2 {\cal F} g^{-1}
  {\cal F} \right]_{ij}.
\end{equation}

In order to find commutators of coordinates and momenta, we use the
relation between the commutators and Dirac brackets as follows
\begin{equation}
  \label{eq:com-D}
  [ z, w ] = i \{ z, w \}_{\rm D},
\end{equation}
where Dirac bracket is given by
\begin{equation}
  \label{eq:D-PB}
  \{ z, w \}_{\rm D} = \{ z, w \} - \{ z, \Omega_i\}
  (\Delta^{-1})^{ij} \{ \Omega_j, w \},
\end{equation}
for $z$ and $w$ any functions of coordinates and momenta.  Using Poisson
brackets of the constraints (\ref{eq:d-PB}), one can construct the 
corresponding inverse matrix   
\begin{eqnarray}
  \label{eq:inv-mat}
  \left( \Delta^{-1} \right)^{ij} &=& \frac{\epsilon^2}{(2\pi\alpha')^2}
      \left[ {\cal F}^{-1} g (G+M)^{-1} \right]^{ij} \nonumber \\
  &=& \frac{\epsilon^2}{(2\pi\alpha')^2} \left[ \left( G+M \right)^{-1} g {\cal F}^{-1} \right]^{ij}.
\end{eqnarray}
to obtain commutation relations as follows
\begin{eqnarray}
  & & [ x_{(0)}^i, x_{(0)}^j ] = - i (2\pi\alpha')^2 \left[ (G+M)^{-1}
      {\cal F} g^{-1} \right]^{ij}, \label{[]0-1} \\
  & & [ x_{(0)}^i, p_{(0)j} ] = \frac{i}{2} \delta^i_j, \label{[]0-2} \\
  & & [ p_{(0)i}, p_{(0)j} ] = \frac{i}{4}\frac{1}{(2\pi\alpha')^2}
      \left[ g {\cal F}^{-1} (G+M) \right]_{ij}, \label{[]0-3} \\
  & & [ x_{(1)}^i, x_{(1)}^j ] = 0, \label{[]1-1} \\
  & & [ x_{(1)}^i, p_{(1)j} ] = i \delta^i_j, \label{[]1-2} \\
  & & [ p_{(1)i}, p_{(1)j} ] = \frac{i}{(2\pi\alpha')^2} \left[ g
      {\cal F}^{-1} M (G+M)^{-1} M \right]_{ij}, \label{[]1-3} \\
  & & [ x_{(0)}^i, p_{(1)j} ] = i {\left[ (G+M)^{-1} M
      \right]^i}_j, \label{[]mix-1} \\
  & & [ p_{(0)i}, p_{(1)j} ] = - \frac{i}{2}\frac{1}{(2\pi\alpha')^2}
      \left( g {\cal F}^{-1} M \right)_{ij}, \label{[]mix-2} \\
  & & [ x_{(a)}^i, p_{(b)j} ] = i \delta^i_j \, \delta_{ab}, \qquad\qquad
      \mbox{for }a,b=2,3,4,\cdots \label{[]bulk} \\
  & & \mbox{others} = 0, \label{other}
\end{eqnarray}
where $g_{ij} = (g^{-1})^{ij} = \delta^i_j$ and we have used the fact that
gauge fields can be written as $A_i = - \frac12 F_{ij} x_{(0)}^j$ for
a constant field strength $F_{ij}$. Eqs.~(\ref{[]0-1}), (\ref{[]0-2}),
and (\ref{[]0-3}) describe the geometry on $D$-brane, while
Eqs.~(\ref{[]1-1}), (\ref{[]1-2}), and (\ref{[]1-3}) describe the
geometry near $D$-brane.  As a result, these commutators give 
rise to the noncommutative geometry in the direction parallel to $D$-brane.  
Here the product operation between functions of $x_{(a)}^i$ is described in 
terms of the Moyal bracket~\cite{sw}, instead of ordinary product. 
Eqs.~(\ref{[]mix-1}) and (\ref{[]mix-2}) reflect the fact that $D$-brane and 
the bulk near $D$-brane interact each other since the commutators of 
variables on the $D$-brane and $p_{(1)j}$ do not commute.  As shown in 
Eq.~(\ref{[]bulk}), the region out of $D$-brane has commutative geometry.

\section{BFT Embedding of Noncommutative $D$-brane system}
\label{sec:BFT}
Since the boundary conditions are second class constraints, the 
noncommutativity of spacetime on $D$-brane arises as shown in the previous
section.  As a result, some kind of local symmetry has been broken due to the 
existence of these second class constraints.  Now let us use the BFT method 
to recover the symmetry which converts the second-class constraints into 
first-class ones.  From now on, we assume that $r=2$, namely $i,j = 1,2$, 
for simplicity.  In order to construct a first-class constraint system from 
the Hamiltonian (\ref{eq:d-Hc}) and the second class constraints
(\ref{eq:d-constr}), we introduce new auxiliary variables $y^i$
satisfying 
\begin{equation}
  \label{eq:d-y}
  \{ y^i, y^j \} = \omega^{ij},
\end{equation}
where $\omega^{ij}$ are antisymmetric constants.  New constraints are then 
defined as
\begin{equation}
  \label{eq:d-constr-}
  \tilde{\Omega}_i = \Omega_i + \gamma_{ij} y^j \approx 0,
\end{equation}
where $\gamma_{ij}$ are constants.  Here $\omega^{ij}$ and $\gamma_{ij}$
will be determined to satisfy the strongly involutive constraint 
algebra $\{ \tilde{\Omega}_i,\tilde{\Omega}_j \} \approx 0$, which yields the 
algebraic relation
\begin{equation}
  \label{eq:d-PB-}
  \Delta_{ij} + \gamma_{ik} \, \omega^{kl} \, \gamma_{jl} = 0.
\end{equation}
Without any loss of generality \cite{fg,kp}, the solutions of 
Eq.~(\ref{eq:d-PB-}) can be solved through the ansatz 
\begin{equation}
  \omega^{ij} = \epsilon^{ij}, \label{eq:d-w} 
\end{equation}
\begin{equation}
\gamma_{ij}=\left(\begin{array}{cc}
 1 & 0\\
 0 & -\Delta_{12}
\end{array}\right).
\label{eq:d-r}
\end{equation}

Using Eqs.~(\ref{eq:d-y}) and (\ref{eq:d-w}) one can define conjugate pair
($y$,$p_y$) by
\begin{eqnarray}
  y &=& y^1, \label{def-y} \\
  p_y &=& y^2, \label{def-py}
\end{eqnarray}
where the new auxiliary variables $y$ and $p_y$ can be interpreted as new 
coordinate and its conjugate momentum, respectively.  Eq.~(\ref{eq:d-y}) then 
describes the ordinary Poisson brackets.  Then, using Eqs. (\ref{eq:d-w}), 
(\ref{eq:d-r}), (\ref{def-y}), and (\ref{def-py}), we obtain the desired 
first class system possessing the modified Hamiltonian, which 
satisfies $\{ \tilde{\Omega}_i, \tilde{H}_c \} =0$, 
\begin{eqnarray}
  \label{eq:m-H}
  \tilde{H}_c &=& H_c - 
      \frac{2\pi\alpha'}{\epsilon^2 \Delta_{12}} y \left[
      M_{22} p_{(1)}^2 - (Mg^{-1}B)_{21} (x_{(2)}^1 - x_{(1)}^1) -
      \theta_{21}^{-1} (x_{(1)}^1 - x_{(0)}^1) \right]\nonumber\\
   & & \quad\ \, - \frac{2\pi\alpha'}{\epsilon^2} p_y \left[ M_{11} p_{(1)}^1 -
      (Mg^{-1}B)_{12} (x_{(2)}^2 - x_{(1)}^2) - \theta_{12}^{-1}
      (x_{(1)}^2 - x_{(0)}^2) \right]\nonumber\\
   & & \quad\ \, + \frac{\pi\alpha'}{\epsilon^3} ( G + Mg^{-1}M^T
      )_{11} (\frac{1}{(\Delta_{12})^2} y^2 + p_y^2),
\end{eqnarray}
and the first constraints, which satisfy $\{ \tilde{\Omega}_i, 
\tilde{\Omega}_{j} \} =0$, 
\begin{eqnarray}
  \label{eq:m-constr}
  \tilde{\Omega}_1 &=& \Omega_1 +  y,
                       \nonumber \\ 
  \tilde{\Omega}_2 &=& \Omega_2 - \Delta_{12} \, p_y.
\end{eqnarray}
Next, in order to obtain the corresponding Lagrangian from the first class 
Hamiltonian (\ref{eq:m-H}), we consider the partition function 
\begin{equation}
  \label{eq:Z}
  Z = \int \prod_{a=0} \prod_{i,j} {\cal D}x_{(a)}^i {\cal D}y 
     {\cal D}p_{(a)j} {\cal D}p_y \det\{ \tilde{\Omega}_i,\Gamma_j \}
     \delta[\Gamma_i] \delta[\tilde{\Omega}_i] \exp \left[ 
     i \int d\tau \left( \sum_{a=0} p_{(a)i} \dot{x}_{(a)}^i + p_y
     \dot{y} - \tilde{H}_c \right) \right],
\end{equation}
where $\Gamma_i = 0$ are gauge fixing conditions which are
chosen as the functions for coordinates only.  Integrating out momenta $p_{(a)i}$ 
and $p_y$, we obtain 
\begin{equation}
  \label{eq:Z-L}
  Z = \int \prod_{a=0} \prod_{i,j} {\cal D}x_{(a)}^i {\cal D}y 
      \det\{ \tilde{\Omega}_i,\Gamma_j \} \delta[\Gamma_i] 
      e^{i \int d\tau \tilde{L}},
\end{equation}
where the effective Lagrangian is given by
\begin{eqnarray}
  \label{eq:m-L}
  \tilde{L} &=& \frac{\epsilon}{4\pi\alpha'} \sum_{a=1} \left\{
    \left( \dot{x}_{(a)}^i\right)^2 + 4\pi\alpha' B_{ij}
    \dot{x}_{(a)}^i \frac{1}{\epsilon} \left( x_{(a+1)}^j - x_{(a)}^j
    \right) - \frac{1}{\epsilon^2} \left( x_{(a+1)}^i - x_{(a)}^i
    \right)^2  \right\} \nonumber \\
  & & + \frac{1}{4\pi\alpha' \epsilon} \left[ g - (2\pi\alpha')^2
    Bg^{-1}B \right]_{ij} \left( x_{(2)}^i - x_{(1)}^i \right) \left(
    x_{(2)}^i - x_{(1)}^i \right) \nonumber \\
  & & - \dot{x}_{(0)}^i \left[ A_i + \frac{1}{(2\pi\alpha')^2}({\cal
    F}^{-1}M)_{ij} \left( x_{(1)}^j - x_{(0)}^j \right) -
    \frac{\epsilon}{(2\pi\alpha')^2} {\cal
    F}_{1i}^{-1} y \right] \nonumber \\
  & & + \frac{M_{22}}{\epsilon\Delta_{12}}
    \dot{x}_{(1)}^2 y -
    \frac{\pi\alpha'}{\epsilon^3(\Delta_{12})^2} G_{11} y^2 + 
    \frac{2\pi\alpha'}{\epsilon^2 \Delta_{12}}
    \theta_{12}^{-1} y \left( x_{(1)}^1 - x_{(0)}^1 \right) \nonumber \\
  & & + \frac{\epsilon^3}{4\pi\alpha'} G_{11}^{-1} \left[
    \dot{y} + \frac{1}{\epsilon} \left\{
    M_{11}\dot{x}_{(1)}^1 -
    \frac{\epsilon^2 \,\Delta_{12}}{(2\pi\alpha')^2} {\cal F}_{21}^{-1}
    \dot{x}_{(0)}^1 - \frac{2\pi\alpha'}{\epsilon} \theta_{12}^{-1}
    \left( x_{(1)}^2 - x_{(0)}^2 \right) \right\} \right]^2,
\end{eqnarray}
with ${\cal F}_{12}^{-1} = - \frac{1}{{\cal F}_{12}}$, $G_{11}^{-1}
= \frac{1}{G_{11}}$, and $\Delta_{12}=
\frac{(2\pi\alpha')^2}{\epsilon^2}(G_{11} + M_{11}) {\cal F}_{12}$.
When we quantize this system, the commutation relations are given by
the usual Poisson's bracket, not by Dirac's one, because the constraints 
have become first-class.  One can then obtain the nonvanishing commutators
\begin{eqnarray}
  & & [ x_{(a)}^i, p_{(b)j} ] = i \delta^i_j \delta_{ab}, \label{xp}
  \\
  & & [ y, p_y ] = i. \label{yp}
\end{eqnarray}
As a result, the extended geometry including the coordinate $y$, becomes 
commutative at the
boundary as well as in the bulk of open string.  The product of functions of 
($x_{(a)}^i$, $y$) is then ordinary one, not the Moyal bracket any more.

On the other hand, since the effective Lagrangian gives first-class
constraints, there exist some additional symmetries. To find the proper 
transformation rule corresponding to these symmetries we use the Hamiltonian
formulation. The primary action with both coordinates and momenta is 
given as 
\begin{equation}
  \label{eq:p-S}
  \tilde{S}_0^p = \int d\tau \left[ \sum_{a=0} p_{(a)i}
  \dot{x}_{(a)}^i + p_y \dot{y} - \tilde{H}_c - \tilde{u}^i
  \tilde{\Omega}_i \right],
\end{equation}
and the transformation generator in the corresponding symmetry group is then 
given by
\begin{equation}
  \label{eq:gen}
  G = \varepsilon^i \tilde{\Omega}_i,
\end{equation}
where $\varepsilon^i = \varepsilon^i(\tau)$ are the local parameters for the
transformation. Since the transformation is given by $\delta z = \{ z, 
G \}$ for any variable $z$, the coordinates and momenta in 
the modified Hamiltonian $\tilde{H}_c$ transform as
\begin{eqnarray}
  \label{eq:transf-x}
  & & \delta x_{(0)}^{i} = -\frac{1}{\epsilon} (2\pi\alpha')^2 {\cal
        F}_{ij} \varepsilon^j, \nonumber \\
  & & \delta x_{(1)}^i = 0, \nonumber \\
  & & \delta x_{(a)}^i = 0, \qquad \mbox{for $a=2,3,\cdots$} \nonumber \\
  & & \delta y = - \Delta_{12}\, \varepsilon^2,
\end{eqnarray}
and
\begin{eqnarray}
  \label{eq:transf-p}
  & & \delta p_{(0)i} = \frac{1}{2\epsilon} \left( G+M\right)_{ij}
      \varepsilon^j
      \nonumber \\
  & & \delta p_{(1)i} = -\frac{1}{\epsilon} M_{ij} \varepsilon^j,
      \nonumber \\
  & & \delta p_{(a)i} = 0, \qquad\mbox{for $a=2,3,\cdots$} \nonumber \\
  & & \delta p_y = - \varepsilon^1.
\end{eqnarray}
As a result, the action (\ref{eq:p-S}) is invariant under the local
transformation Eq.~(\ref{eq:transf-x}).  Now it 
seems appropriate to comment on the Lagrangian multipliers $\tilde{u}^{i}$.  
The transformations of these multipliers are determined to keep the action 
(\ref{eq:p-S}) invariant under the transformations (\ref{eq:transf-x}) and 
(\ref{eq:transf-p}).   Since the number of the transformation parameters is 
two, we may choose gauge fixing conditions as $y=0$ and $p_y=0$ 
simultaneously.  For the above gauge fixing conditions the multipliers
$\tilde{u}^i$ are determined by 
\begin{eqnarray}
  \tilde{u}^1 &=& \frac{2\pi\alpha'}{\epsilon^2 \Delta_{12}}
      \left[ M_{22} p_{(1)}^2 - (Mg^{-1}B)_{21} (x_{(2)}^1 - x_{(1)}^1) -
      \theta_{21}^{-1} (x_{(1)}^1 - x_{(0)}^1) \right], \label{u1}\\
  \tilde{u}^2 &=& -\frac{2\pi\alpha'}{\epsilon^2 \Delta_{12}} 
      \left[ M_{11} p_{(1)}^1 - (Mg^{-1}B)_{12} (x_{(2)}^2 -
      x_{(1)}^2) - \theta_{12}^{-1} (x_{(1)}^2 - x_{(0)}^2)
      \right] \label{u2}
\end{eqnarray}
because the primary Hamiltonian is given by $\tilde{H}_{p}=\tilde{H}_{c}+
\tilde{u}^{i}\tilde{\Omega}_{i}$.  Note that imposing these gauge fixing 
conditions the modified Hamiltonian $\tilde{H}_c$ is exactly reduced to the 
original Hamiltonian $H_c$ of the open string. 

Finally, applying the solutions of the equations of motion for the momenta 
and multipliers to the primary action (\ref{eq:p-S}), we obtain the desired
effective action, which is invariant under the local transformation 
Eq.~(\ref{eq:transf-x}), in terms of coordinates only
\begin{equation}
  \label{eq:m-S}
  \tilde{S} = \int d\tau \tilde{L},
\end{equation}
where $\tilde{L}$ agrees with the effective Lagrangian (\ref{eq:m-L}).

\section{Conclusions}
\label{sec:dis}

In conclusion, we have newly found that the coordinates of $D2$-brane with a 
constant Neveu-Schwarz $B$-field become commutative in the BFT quantization 
scheme.  Furthermore, we have shown that in these systems, the Moyal bracket 
also becomes the ordinary product since the geometry is commutative.  However 
these properties have originated from discretized action, not from continuous 
action.  Therefore, it is interesting to apply the BFT method to highly 
nontrivial continuous system of string theory through further rigorous 
investigation, although one can conjecture that the BFT embedded systems in 
both discrete and continuous actions have the symmetry corresponding
to the first-class constraints system.

\section*{Acknowledgments}
This work was supported by Ministry of Education, BK 21, Project
No. D-0055, 1999.

\end{document}